Title:  Observation of Cholesterol Dissolved in Microscopic Deposits of Free Fatty Acids in the Lumen of an Aorta of a Mouse Model for Human Atherosclerosis
Author:  Fran Adar, Ph.D.  (Horiba Jobin Yvon, Edison, NJ
Comments: 5 pages, 3 figures

The Raman spectrum of microscopic droplets of lipid material on the lumen of the aorta of an apolipoprotein E knock-out mouse were reported in the proceedings of an SPIE conference.  Based on the absence of the carbonyl band in the spectrum, at that time it was determined that the spectrum represents free fatty acids rather than triglycerides.  More recent examination of the spectrum indicates that these droplets contain dissolved cholesterol, and thus can be used as an early indicator of the atherosclerosis process in animal models during drug development.

**Introduction**
Lipids, including cholesterol, triglyicerides, and lipoproteins have been implicated in the development of atherosclerosis[1].  Typically, blood tests evaluate the various lipoproteins (high density HDL, low density LDL, and very low density VLDL) as well as triglycerides that circulate in the blood serum.  What is difficult to asses is the chemistry occurring locally on the arterial wall before and during the development of atherosclerotic plaque.  And, in addition, it is difficult to differentiate between vulnerable plaque and those lesions that are less likely to produce cardiac events.  Raman spectroscopy has the advantage of being able to characterize the molecular composition of arterial vessel changes and than be correlated with plaque vulnerability.

A murine model for human atherosclerosis has been developed[2].  These animals, lacking in apolipoprotein E (the proteinaceous component of HDL), are termed apoE knock-out mice.  The aorta from one such mouse, was examined *ex situ*, without fixing or staining, in a Raman microscope.  Results from those measurements were published in the SPIE proceedings of the meeting in which the results were presented[3].  At the time, three observations were summarized:
1. Calcified regions showed evidence for calcite, a particular crystalline form of $CaCO_3$.
2. The amount of unsaturation can be tracked in the intensity of the >C=C< bond versus the intensity of the >$CH_2$ deformations.
3. The presence of globular deposits of the order of 3 to 8 µm was observed, and the spectrum indicated that these deposits were composed of free fatty acids rather than esterified fats, as would be expected.

Other workers have reported and discussed the molecular origins of the Raman spectra of atherosclerotic plaque.  In 1992 Michael Feld[4] made the earliest measurements of atherosclerotic plaque of which we are aware.  There is a thorough accounting of the spectral features of the major chemical species found in the various regions of the vessels examined.  However, the spatial resolution of the original system used was 1mm which precluded identifying microscopic accretions. In a later publication[5] the same group continued this work in order to develop a correlation between chemical composition and morphology, and plaque instability, in order to predict disease progression. They state, however, that (p62) "the spectral variation for any one morphologic structure is very small, indicating that the chemical composition of any morphologic structure is relatively constant." Inspection of the spectra in their Figure 2 shows that the adventitial adipocyte has a clear band near 1740 cm$^{-1}$ representing the esterified carbonyl, whereas foam cells and necrotic cores is lacking this band, but no comment is made regarding the significance of free fatty acids. The presence of cholesterol, however, was clearly recognized.

The research group of Gerwin Puppels at Erasmus University in Rotterdam has also been active in this area. In 1992 they published Raman maps of ceroid (lipid peroxidation products) within an atherosclerotic lesion [6] produced by principle component analysis and then K-means clustering analysis in which 5 clusters was found to account for the variance in the data set. The spectra of 13 components were considered in reconstructing the data set, but "the free fatty acids and triolein have been grouped, and are indicated as triglycerides (TG)" (legend to Figure 2).

In this note, it will be shown that the spectrum that was generated from the globule in the lumen of the apo E knock-out mouse represents a mixture of free fatty acid and cholesterol. While these compounds have been recognized in earlier publications, their presence to the near exclusion of anything else in the small globule deposits noted here is new.

**Results**

Figure 1 is reproduced from the SPIE publication. It shows the spectrum of the globule (middle) in comparison to spectra of fatty tissue (top) and proteinaceous material (bottom). In that publication, the spectrum of the globule was attributed to free fatty acid and not esterified fatty acids based on the absence of the >C=O band at about 1735 cm$^{-1}$. [This is based on the presence either of dimers of the fatty acids, or the carboxylate ion; neither form will exhibit a carbonyl near near 1735 cm$^{-1}$.][7] The additional sharp bands, especially the one at ca. 700 cm$^{-1}$ was ignored at the time. It was recently recognized that this spectrum could represent the spectrum of cholesterol dissolved in free fatty acid, so a model measurement was made in which cholesterol was dissolved in oleic acid, and its Raman spectrum is reported here.

Figure 2a shows the spectrum of cholesterol dissolved in oleic acid (middle) compared to the spectrum of pure oleic acid (top) and cholesterol crystals (bottom). Figure 2b shows the difference spectrum of the solution minus the oleic acid solvent vs. the cholesterol crystals. It is clear that all of the cholesterol bands appear in the difference spectrum, even the band of the carbon double bond at about 1680 cm$^{-1}$ which is buried under the double bond of the oleic acid.

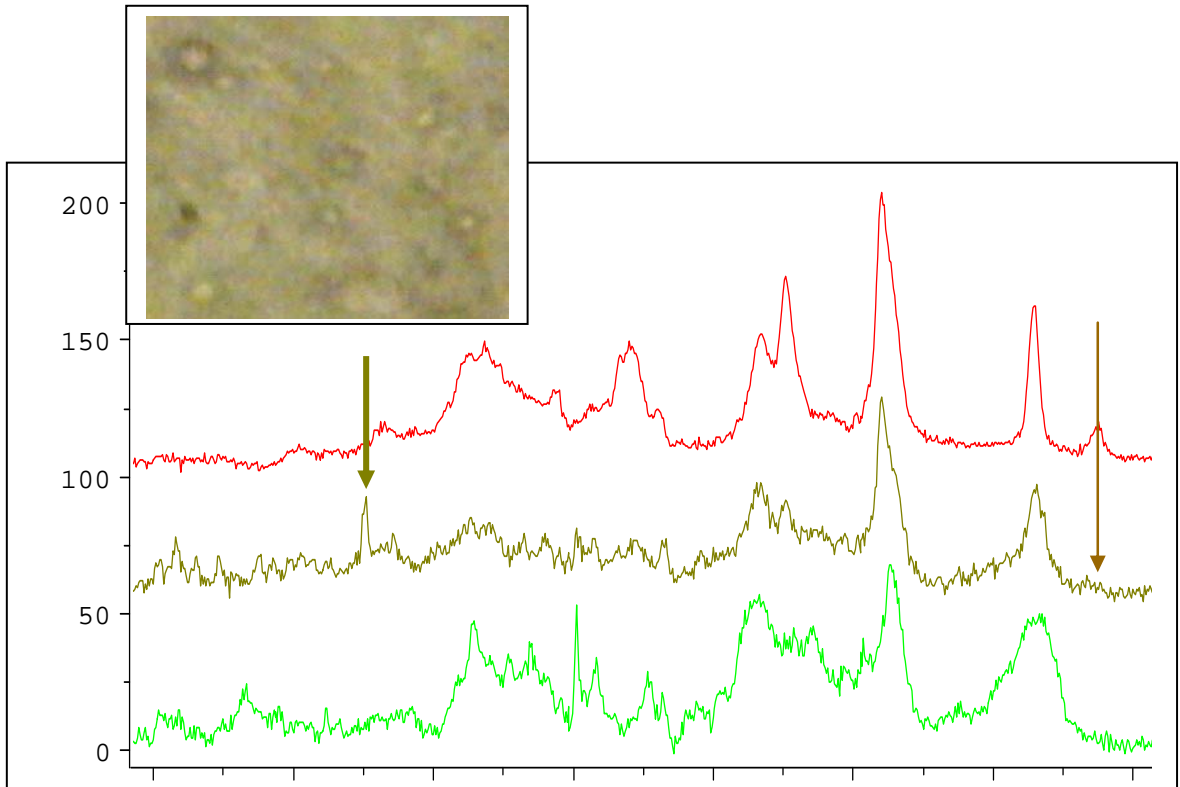

Figure 1. Micrograph of lumen side of aorta of knock-out mouse exhibiting globule deposits (top left). Raman spectra of one of these deposits (middle) vs. a spectrum from a fatty region (top), and s spectrum from a proteinaceous-rich region (bottom).

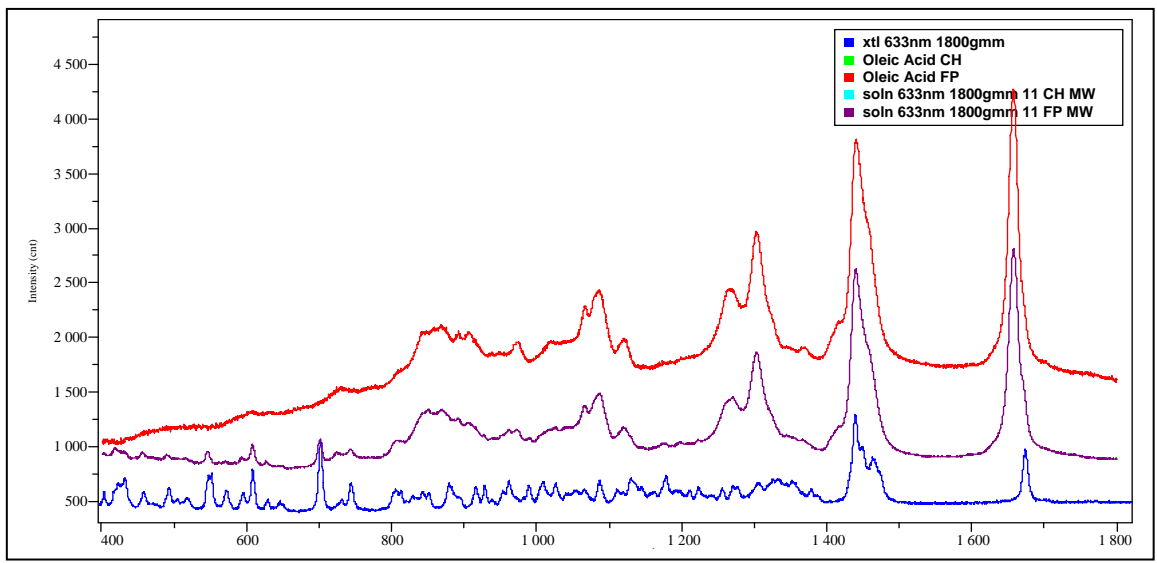

Figure 2a. Raman spectrum of oleic acid (top), cholesterol crystals (bottom) and cholesterol dissolved in oleic acid (middle).

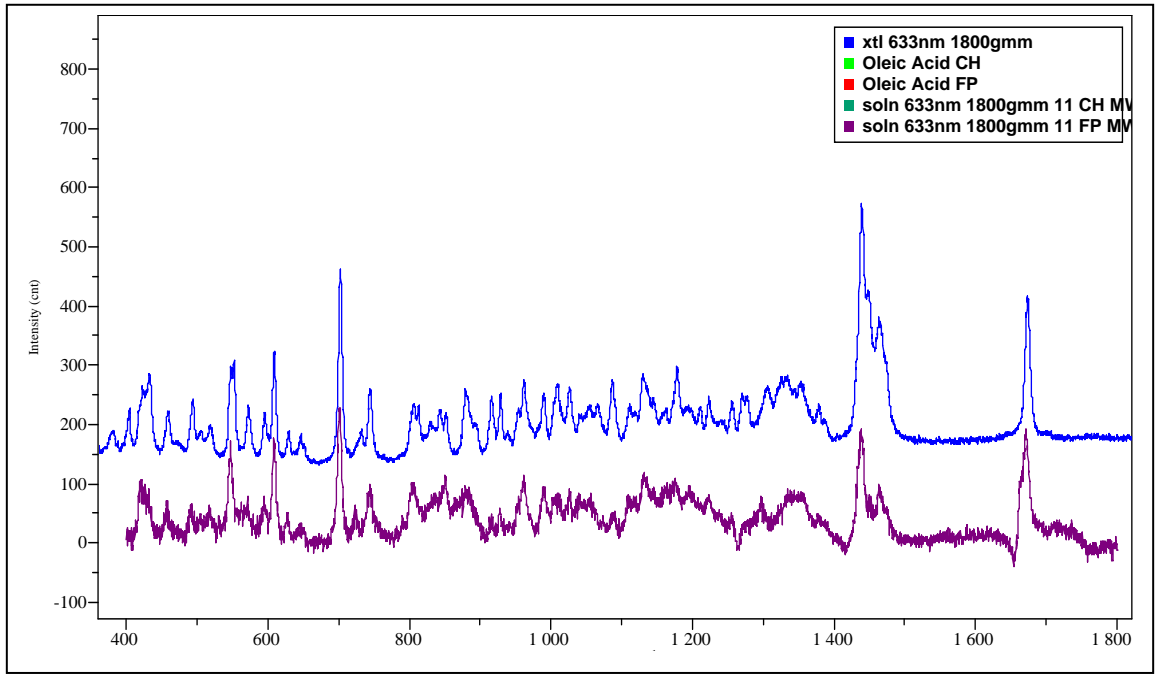

Figure 2b. Spectrum of cholesterol crystals (top) vs. spectrum of cholesterol-oleic acid solution, after subtracting the contributions from the oleic acid.

Finally in Figure 3, the spectra in Figure 1 are overlaid with the spectra of Figure 2a to illustrate the good match between the solution spectrum with that of the fat globule in the aorta of the knock-out mouse.  These spectra support the conclusion that the globule deposits are composed of unsaturated free fatty acids with dissolved cholesterol.

## Summary and Conclusions

Deposits of liquid-like globules were observed on the surface of the lumen side of an apo E knock-out mouse model for atherosclerosis.  Based on standard interpretation of vibrational spectra, and spectra of a model solution of oleic acid with dissolved cholesterol, these deposits are identified as free fatty acids with dissolved cholesterol.

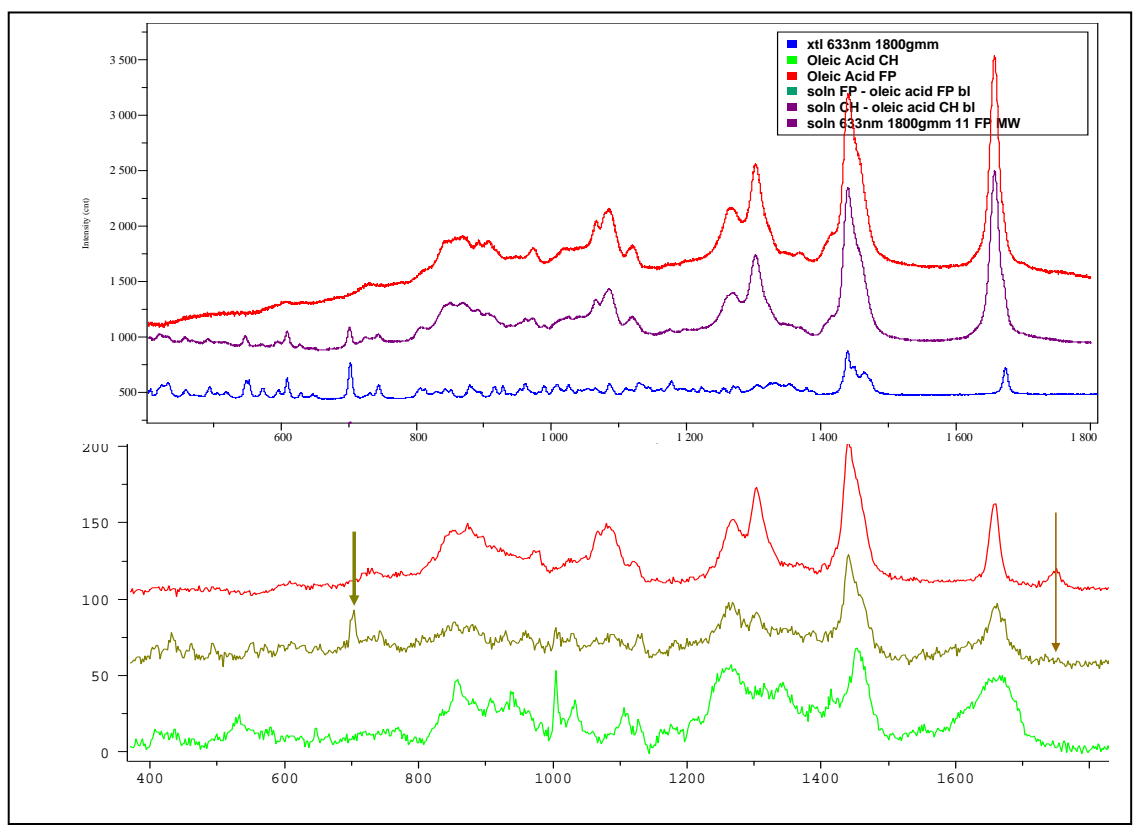

Figure 3. Raman spectra of the globule on the luminal side of the apo E knock-out mouse (middle spectrum of bottom part of figure) compared to the spectrum of cholesterol dissolved in oleic acid (middle spectrum of top of figure).